\documentclass[%
 reprint,
 amsmath,amssymb,
 aps,
]{revtex4-2}

\usepackage{graphicx}
\usepackage{dcolumn}
\usepackage{bm}
\usepackage{hyperref}
\usepackage{bbm}

\newcommand{\moly}{\text{MoTe}_2}

\newcommand{\te}{\text{e}}
\newcommand{\tu}{\text{u}}

\newcommand{\tT}{\text{T}}

\newcommand{\vv}{\boldsymbol{v}}
\newcommand{\vx}{\boldsymbol{x}}

\newcommand{\vV}{\boldsymbol{V}}

\newcommand{\vmu}{\boldsymbol{\mu}}
\newcommand{\vtheta}{\boldsymbol{\theta}}
\newcommand{\vnu}{\boldsymbol{\nu}}

\newcommand{\cA}{\mathcal{A}}

\newcommand{\cE}{\mathcal{E}}

\newcommand{\cI}{\mathcal{I}}

\newcommand{\cS}{\mathcal{S}}

\newcommand{\beq}{\begin{equation}}
\newcommand{\eeq}{\end{equation}}
\newcommand{\ba}{\begin{aligned}}
\newcommand{\ea}{\end{aligned}}
\newcommand{\bc}{\begin{cases}}
\newcommand{\ec}{\end{cases}}
\newcommand{\bmx}{\begin{pmatrix}}
\newcommand{\emx}{\end{pmatrix}}
\newcommand{\la}{\label}

\newcommand{\bfig}{\begin{figure}}
\newcommand{\efig}{\end{figure}}
\newcommand{\ing}{\includegraphics}

\begin{document}

\preprint{APS/123-QED}

\title{Dynamic zoom-in detection of exfoliated two-dimensional crystals using deep reinforcement learning}

\author{Stephan Kim}
\affiliation{Department of Electrical and Computer Engineering, Princeton University, NJ 08544}

\begin{abstract}
Owing to their tunability and versatility, the two-dimensional materials are an excellent platform to conduct a variety of experiments. However, laborious device fabrication procedures remain as a major experimental challenge. One bottleneck is searching small target crystals from a large number of exfoliated crystals that greatly vary in shapes and sizes. We present a method, based on a combination of deep reinforcement learning and object detection, to accurately and efficiently discover target crystals from a high resolution image containing many microflakes. The proposed method dynamically zooms in to the region of interest and inspects it with a fine detector. Our method can be customized for searching various types of crystals with a modest computation power. We show that our method outperformed a simple baseline in detection tasks. Finally, we analyze the efficiency of the deep reinforcement learning agent in searching crystals. Codes are available at \url{https://github.com/stephandkim/detect_crystals}.
\end{abstract}

\maketitle


\section{Introduction}

    \subsection{Challenges in two-dimensional materials}
    
        The two-dimensional (2D) materials are of great interest in many frontiers of research, from fundamental physics to device engineering. These exfoliable materials can be prepared and studied in a variety of forms, such as monolayers \cite{graphene}, thin films \cite{mote2_edge_supercurrent, eavesdropping}, and bulk crystals \cite{mote2_bulk}. In the monolayer limit, one can stack or twist \cite{twisted} monolayers of same or different materials, using advanced nanofabrication techniques. The tunability and versatility of these materials make them an ideal platform to study emergent physics, create new electronics, and explore new interfaces.
        
        The laborious device fabrication procedures of 2D materials remains as one of the major experimental challenges. To create a device, one first exfoliates the 2D material of interest on to a substrate. The exfoliated microflakes are located randomly and can range from a few to hundreds in number, as shown in Fig. \ref{fig:device_images}. They also vary enormously in size, dimensions, and qualities. The researcher examines every single one of the exfoliated crystals under a microscope until adequate crystals that meet the desired conditions are found. Searching target crystals is one of the bottlenecks in the device fabrication procedures, and it can take up to days to find one good candidate crystal.

    \subsection{Previous works and problems}
    
        Searching target crystal is analogous to the object detection task in computer vision; it consists of two steps, which are localizing and identifying objects (target crystals in this case). Inspired by the similarities, researchers have employed various machine learning techniques to assist the searching process. Shin et al. \cite{2d_mat} combined a motorized microscope with deep neural networks to detect target graphene and boron nitride microflakes. Han et al. \cite{adv_mat} used deep learning architectures to identify target crystals and their thicknesses. Masubuchi et al. \cite{npj} classified crystals based on optical images through data-driven machine learning techniques. Saito et al. \cite{u_net} developed algorithms based on U-Net to determine the thicknesses of exfoliated crystals from optical images. Masubuchi et al. \cite{robot} married computer vision techniques with a robotic system to search target flakes.

        Previous works have mostly focused on identifying target crystals, leaving the localization part of searching problem unsolved. The aforementioned methods identify or detect target crystals from zoomed-in images as Fig. \ref{fig:device_images} (b), which contain only a few crystals. A substrate with exfoliated crystals typically contains a much larger number of crystals as shown in Fig. \ref{fig:device_images} (a). As a result, one has to first locate potential target crystals and obtain zoomed-in images of them, prior to identifying them through the proposed methods. Such a process requires further efforts. Previous studies relied on additional setups, such as motorized microscopes, robotic systems, or manual collection, which themselves become challenges for installation and operation.
        
        A simple solution to resolving both the location and identification issues in the searching problem is using the standard object detection techniques on high resolution images as Fig. \ref{fig:device_images}. However, such an approach, too, is problematic, because of several reasons. The dimensions of crystals are very small, and the extracted features vanish as they are passed on to the following convolutional neural network (CNN) layers \cite{sod1, sod2}. Possible solutions for small object detection, such as tuning the neural network structures, are often cumbersome and difficult to implement. One can further increase the resolution of image, so that crystals are described by a larger number of pixels. However, this increases the computational cost tremendously. An improved yet simple solution is determining a small region from a high resolution image, which is likely to contain target crystals, and analyzing that region with a fine detector.

    \subsection{Dynamic zoom-in}
    
        Here, we propose the method to efficiently and accurately detect target crystals from a high resolution image that contains a large number of crystals by \textit{dynamically zooming in} to the region of interest (ROI). Our method consists of two stages: coarse and fine search. In the coarse search, a deep reinforcement learning (DRL) agent \cite{sutton} determines the ROI based on the information from a downsampled and filtered image. The DRL agent navigates the ROI in different directions until it encloses potential target crystals. Then, a fine detector analyzes the ROI in the full-size image for target crystals. 
        
        Although the idea of zooming-in for efficient object detection has already been explored in the computer vision community \cite{gao,uzkent}, our method differs from them in various aspects. In these previous works, the environments of reinforcement learning (RL) framework are full RGB colored images, and the state spaces of DRL agent are comprised of feature vectors extracted by CNN layers on those images. Pixel-based RL is known for its sample-inefficiency \cite{pixel1, pixel2}. Instead, our method is state-based by utilizing the unique image setting of exfoliated crystals on substrates. The resulting RL environment is a highly compressed binary image. The state and action spaces consist of few vectors, where its dimensions are image-agnostic. Together, these settings require a very compact neural network, use modest computation resources, and show quick convergence during training the DRL agent \cite{sm}.

        The method presented in this work is a high-level design and can be easily customized for searching different types of exfoliated crystals. It is highly modular in that the coarse search is decoupled from the fine search; the ROI proposal from the DRL agent is not based on the information from fine search. As a result, the DRL agent works for any images, so long as they mainly consist of exfoliated 2D materials and a substrate. If one wishes to detect different crystals from previous experiments, he or she can simply replace the fine detector with another and continue the search with the same DRL agent. As a demonstration, we implement our method to search target MoTe$_2$ crystals, which are suitable for studying edge supercurrents \cite{mote2_edge_supercurrent, eavesdropping}, from a high resolution image. In this setting, target crystals are in the thin film limit with tens of layers and have sharp physical edges.

\section{Proposed method for detecting target crytals}

    \subsection{High resolution image and RL environment}
        A full-size, high resolution image $\cI$ that contains all crystals as Fig. \ref{fig:device_images} (a) is downsampled and filtered to create an environment $\cE$, shown in Fig. \ref{fig:architecture} (a), for the DRL agent in our method. The dimensions of $\cI$ are $(H, W)$, where $H$ is the number of pixels along the y-axis and $W$ that along the x-axis. The dimensions of $\cE$ are $(h, w)$. The environment is a binary image, and any objects other than the background are represented by the pixels with the value of one. These objects have the potential to be target crystals when they are scanned under a fine detector. We dub such an object as a \textit{polygon} $p$ hereafter to distinguish them from target crystals.
        
        A series of procedures transforms $\cI$ into $\cE$: k-means clustering, max pooling, downsampling, and size filtering as shown in Fig. \ref{fig:architecture}. Any information in the environment, besides the locations of polygons, is unnecessary. Thus, we reduce the size of original image by taking advantage of the nearly bimodal color distribution in it. For a substrate with exfoliated flakes, the two most dominant colors are that of substrate (background) and that of crystals. The colors of crystals may vary, but such a variation is highly concentrated when it is compared to the color of background. In this setting, the standard clustering techniques can accurately distinguish the pixels of any objects from those of the background. The max pooling operation is then performed on the image to keep the locations of small polygons that are otherwise lost. The downsampling procedure further reduces the size of the image. Finally, a size filter removes polygons that are way too large to be target crystals. The threshold for this filter is determined by the user. The resulting $\cE$ contains $N$ polygons. Based on the standard flood fill algorithm, all polygons $\{p\}$ and their locations are identified prior to search.

    \subsection{DRL agent: coarse search}
        Figure \ref{fig:workflow} shows the workflow of our method. The DRL agent proposes the ROI based on $\cE$. The fine detector scrutinizes the corresponding region in $\cI$. The detected target crystals are recorded and the remaining $p$ in $\cE$ are updated. One iteration of this process is an episode $E$, and episodes repeat until no $p$ is left in $\cE$.
        
        We formulate the process of ROI proposal as a RL problem \cite{sutton}. Figure \ref{fig:architecture} shows the reinforcement learning diagram. At each step $t \in \{0, \cdots, T\}$, the DRL agent takes an action $a \in \mathcal{A}$ to \textit{move} the ROI that maximizes the reward $r_t$, based on the observation of current state $s_t \in \mathcal{S}$. An episode $E$ consists of such steps, and it terminates when the agent voluntarily stops $E$ or $t$ reaches the maximum step $t=T$. Upon termination, the DRL agent gets rewarded $r_T$ based on the polygons that lie within the ROI. The polygons that are fully enclosed in the ROI are removed from $\cE$.
        
        The ROI has a shape of a box in $\cE$ and it is described by by $b = (\vv, h_b, w_b)$, where $\vv$ is the box vector that points to the center of box, $h_b$ its height and $w_b$ its width, respectively. The box moves every turn, hence $\vv_t$ at $t$. A pixel in $\cE$ is described as a vector $\vx$. A polygon $p$ consists of contiguous pixels $\{\vx\}$. Its location is expressed by its center of mass is $\vmu$. If all $\vx$ of a $p$ are within $b$, such $p$ is referred as an enclosed polygon $p_\te$. Others are called unenclosed polygons $p_\tu$. Figure \ref{fig:architecture} (b) shows examples of $p_\te$ and $p_\tu$. The DRL agent can perceive up to $M$ enclosed polygons in $b$. This value is set during training. Because $\vv$ changes every $t$, the types of $p$ are subject to change. For instance, $p_\te$ at $t-1$ can become $p_\tu$ at $t$, because one of $\{\vx\}$ goes out of $b$. The coordinates of objects in $\cE$ can be mapped on to those in $\cI$. The box $b$ in $\cE$ corresponds to $B$ in $\cI$ and $B = (\vV, H_B, W_B)$. Again, $\vV$ points to the center of $B$, and $H_B$ and $W_B$ are height and width of $B$, respectively. 
        
        The state space $\mathcal{S}$ consists of a scalar and three vectors, which are $n_\te$, $\vtheta_\te$, $\vtheta_\tu$, and $\vnu$, respectively. The variable $n_\te$ counts the number of $p_\te$ at the current $t$. It is used to calculate the termination reward after the current $E$ finishes. To detect nearby polygons, the DRL agent segments the surroundings of $b$ into eight sections with an increasing angle as Fig. \ref{fig:architecture} (b). The observation vector for enclosed polygons $\vtheta_\te$ encodes the locations of $p_\te$. It has nine components $\vtheta_\te = (\theta_{\te, 0}, \cdots, \theta_{\te, 8})$. The first eight components are related to each of the eight sections. Each of them counts $p_\te$ in its corresponding section. The last component of $\vtheta_\te$ is reserved for the case when $\vmu_\te$ of $p_\te$ is $\vv = \vmu_\te$. When $\vv = \vmu_\te$, $p_\te$ lies on the boundaries of all sections and $(\theta_{\te, 0}, \cdots, \theta_{\te, 7})$ are ill-defined. Then, $\vtheta_{\te, 8}$ becomes nontrivial to resolve this issue. For implementation, $n_e$ and $\vtheta_\te$ are normalized by $M$. The other observation vector $\vtheta_\tu$ contains the information related to the \textit{closest} unenclosed polygon $p_{\tu, 0}$. It is similar to $\vtheta_e$ but has only eight components $\vtheta_\tu = (\theta_{\tu, 0}, \cdots, \theta_{\tu, 7})$, since $\vmu_\tu$ of $p_\tu$ is never $\vmu_\tu = \vv$. The number of $p_{\tu, 0}$ is strictly one, and therefore, $\vtheta_\tu$ is a one-hot vector. The vector $\vv$ changes every $t$ and so does the corresponding $p_{\tu, 0}$. To ensure that $b$ does not go out of bound, the edge vector $\vnu$ is sensitive to the edges of $b$. It is binary and has four components that correspond to the respective four edges of $b$. When one edge reaches the boundary of $\cE$, the corresponding component becomes nontrivial. 
        
        The set of actions $\cA$ consists of nine different $a$ as shown in Fig. \ref{fig:architecture} (c): increasing $\vv$ by one pixel in eight different directions and the stop action. The first eight actions change $\vv$ by one pixel in the respective eight directions. The last terminates the current $E$.

        There are two kinds of rewards: in-episode reward $r_t$ at $t$ and termination reward $r_\tT$ at termination step $T$. At $T$, the DRL agent is rewarded based on $n_{\te}$, which is analogous to a multi-objective reinforcement learning (MORL) problem \cite{MORL_review}. However, our situation is much simpler than the general case because the objectives in our setting are low-dimensional and can be defined by $n_e$. We use a scalar reward as a weighted linear combination of $n_e$ \cite{MORL_linear}. It is \beq
        r_T =  \sum_{k=1}^{M} \frac{k}{M} \mathbbm{1}_{n_e = k} + \mathbbm{1}_{n_e > M},
        \eeq

        \noindent which simplifies to
        \beq
            r_T = \min \Big( \frac{n_e}{M}, 1\Big).
        \eeq

        The in-episode reward $r_t$ is designed such that the DRL agent is gravitated towards maximizing the number of $p_\te$ during $E$. Polygons in the environment are dispersed over a wide range, which can cause the well-known sparse reward problem. To cope with this issue, we choose a potential-based reward. Before describing $r_t$, we first quantify the optimality of $\vv_t$ with respect to the set of enclosed polygons $\{p_\te\}$ at $t$ by the enclosed polygon potential $U^\te_t$. It is defined as 
        
        \beq
            U^\te_t =
            \bc
            \sum\limits_{i}^{n_\te} \frac{1}{\alpha \|\vmu_{\te, i} - \vv\|_2 + 1}, & \text{if } n_\te \geq 1, \\[10pt]
            1/K, & \text{otherwise,}
            \ec
            \la{eq:enclosed_polygon_potential}
        \eeq
        
        \noindent where $\alpha$ is a scaling factor, $K$ a large constant, $K \gg 1$, and $\vmu_{\te, i}$ is $\vmu$ for $i$th $p_\te$. Likewise, the potential for the closest unenclosed polygon $U^\tu_t$ is
        
        \beq
            U^\tu_t = 
            \bc
            \frac{1}{\alpha \|\vmu_{\tu, 0}-\vv_p\|_2 + 1}, & \text{if } p_{\tu, 0}, \\[10pt]
            K, & \text{otherwise.}
            \ec
            \la{eq:unenclosed_polygon_potential}
        \eeq
        
        \noindent The potential $U^\tu_t$ is similar to $U^\te_t$ but the $K$ term differs. The overall potential $U_t$ at $t$ has a form of harmonic mean. It is
        
        \beq
            U_t = \frac{1}{1/U^\te_t + 1/U^\tu_t}.
        \eeq
        
        \noindent The intuition behind $U_t$ is that it encourages the DRL agent to keep all of current $\{p_{\te}\}$ while reaching for $p_{\tu,0}$ as much as possible. The potential $U_t$ maximizes when \textit{both} $U^\te_t$ and $U^\tu_t$ maximize. In addition, the terms involving $K$ in \eqref{eq:enclosed_polygon_potential} and \eqref{eq:unenclosed_polygon_potential} have penalizing and null effects on $U^\te_t$ and $U^\tu_t$, respectively. If the DRL agent fails to secure any $p_\te$, $U^\te_t$ decreases, leading to a small $U_t$. When the detection task nears the completion, $p_{\tu,0}$ does not exist, and $1/U^\tu_t$ vanishes with $U^\tu_t=K$.
        
        The in-episode reward $r_t$ is
        \beq
            r_t =
            \bc
            -1, & \text{if} \ \| \vnu \|_2 > 0, \\
            r_0 (1 - t/T), & \text{if} \ \Delta U_t > 0, \\
            -r_0, & \text{otherwise,}
            \ec
        \eeq

        \noindent where $\Delta U_t$ is the \textit{difference} of potential between the current and previous steps, $\Delta U_t = U_{t} - U_{t-1}$. We utilize the difference of potential, so that $r_t$ agnostic of $N$ in $\cE$. The decaying reward enforces the DRL agent to take the minimum steps possible to maximize the reward hence arriving at $p$ through the fastest route possible. Running a fine detector in $B$ that is out of $\cI$ results in errors. To avoid this, the DRL agent is heavily penalized for pushing $b$ out of bound. In addition, whether $b$ is within $\cE$ is the first condition that is considered when calculating $r_t$. The constant $r_0$ is used to tune $r_t$ and $r_T$, such that securing at least one $p_\te$ remains as the prime reward in our RL framework,
        \beq
            r_T|_{n_\te=1} > \sum_{t=0}^T r_0 (1 - t/T).
        \eeq

        \noindent Therefore,
        \beq
            r_0 = \frac{2}{MT}.
        \eeq
        
        The core idea behind the DRL agent in our method is that it learns the correlations between the locations of $p_\te$, $p_{\tu, 0}$ (represented by $\vtheta_\te$ and $\vtheta_\tu$ in $\cS$) and $\cA$. Due to the compact $\cS$ and $\cA$, the DRL agent does not require a complex neural network to learn the correlations. Moreover, the small neural network is easy and quick to train. Finally, the correlations are strictly based on polygons, which are unrelated to any features extracted by CNN layers in fine detectors. As a result, the DRL agent can be used with different fine detectors, depending on the need.

    \subsection{Fine detector}
        
        The fine detector scans $B$ in $\cI$. Therefore, it is trained on full RGB images that have the dimensions of $B$, $(H_B, W_B)$. The training and detection processes follow the standard object detection procedures. Furthermore, the architectures and detectors in the previous works \cite{adv_mat, 2d_mat, npj, robot, u_net} can be used as the fine detector in our method.

\section{Experiment}

    \subsection{Target $\moly$ crystals for edge superrcurrent measurements}

        
        We apply our method to search target $\moly$ crystals for edge supercurrent measurements \cite{mote2_edge_supercurrent, eavesdropping}. We trained a DRL agent and a fine detector on the previous collected images of target $\moly$ crystals. The performance of our method is compared to a baseline detector. We also analyze the efficiency of different DRL agents for coarse search.
        
        There are several criteria for a target $\moly$ crystal. A target crystal is in the thin film limit, consisting of tens of two-dimensional layers. The thickness $d$ is around $d \sim 100$ nm. Furthermore, there exists an upper bound for the size of target crystals. The edge supercurrent gives rise to periodic modulation of critical current with respect to the applied field. The period of modulation $B_p$ should be well above the resolution of applied magnetic field, and it is $B_p \geq 10 \ \mu$T empirically. Due to the flux quantization, the area of target crystal $A$ is inversely proportional to $B_p$ via $\phi_0 = B_p \cdot A$, where $\phi_0$ is the magnetic flux quantum. The resulting condition for $A$ is $A \leq 200\mu$m$^2$. Finally, a target crystal should have well-defined physical edges for edge transport.

        The dataset contains 230 high resolution images $\{\cI\}$ with the total of 10,360 target crystals. The dimensions of $\cI$, $\cE$, $B$, and $b$ are $(1920, 2448), (120, 153), (320, 320)$, and $(20, 20)$, respectively. The dimensions for all target crystals in $\cI$ are less than 64 pixels in height and width. 

    \subsection{Implementation details}

        We used OpenAI Gym \cite{gym} to implement the RL framework and Stable Baselines 3 package \cite{sb3} to train the DRL agent on PyTorch. The Proximal Policy Optimization (PPO) algorithms \cite{ppo} were used for training. A policy gradient method is usually less sample efficient than a value-based approach is, but it tends to be more stable during training. Because of the simple $\cE$, $\cS$, and $\cA$ in our method, the sample efficiency was of no concern. The neural networks used for actor and critic consisted of two hidden layers of 16 neurons. Four different DRL agents were trained with $M = \{1, 2, 3, 4\}$. The values of $\alpha, K, T$ are set to $\alpha = 0.2, K = 100, T = 200$, respectively.
        
        We used RetinaNet \cite{retinanet} for the fine detector and the baseline detector. Other settings, such as anchors, were identical to those in the original report, if not specified. The optimizer was the standard stochastic gradient descent in PyTorch package with the learning rate of 0.0005 and momentum of 0.9. The fine detector was trained on randomly cropped images with dimensions $(H_p, W_p)$ of target crystals. The baseline detector was trained on augmented $\{I\}$. The training and experiments were conducted using one RTX 3080 Ti GPU. For performance evaluation, tests were conducted for k-fold cross validation with $k=5$. Details of training are in \cite{sm}.

\section{Results}

   \subsection{Qualitative evaluation}
        We analyze the performances of our method and the baseline qualitatively. Figure \ref{fig:qualitative_analysis} shows two results. Additional results are in \cite{sm}. Our method surpassed the baseline in the detection task. Not only did it find a larger number of crystals, but it also detected each of them more accurately than the baseline did. The baseline created redundant bounding boxes for the same crystals, while our method did only one for each.

    \subsection{Quantitative evaluation}
    
        The average precision (AP) and f1 scores are commonly used to measure the efficiency of object detection algorithms. When a detector discovers a potential target crystal, it creates a bounding box around it. The accuracy of this bounding box is determined by the intersection over union (IoU), which is 
        \beq
            \text{IoU} = \frac{\text{Ground Truth} \cap \text{Prediction}}{\text{Ground Truth} \cup \text{Prediction}}.
        \eeq
        
        \noindent The threshold for true positive was set to $0.5$. Precision and recall are then
        \beq
            \ba
                & \text{Precision} = \frac{\text{TP}}{\text{TP+FP}},\\
                & \text{Recall} = \frac{\text{TP}}{\text{TP+FN}},
            \ea
        \eeq
        \textbf{}
        \noindent where TP, FP, and FN are true positive, false positive, and false negative, respectively. The AP score is obtained by integrating the area under the curve in the precision versus recall plot. The f1 score is a harmonic mean of precision and recall, $\text{F1} = 1/(1/\text{Precision} + 1/\text{Recall})$.
    
        We examine the performance of our method using these metrics and compare it to that of the baseline detector in Table \ref{tab:metrics}. The precision and recall values were collected for all $k=5$ cross validation test sets. The results were then concatenated and the comprehensive AP and f1 scores were calculated. Our method exhibited large increases in the AP and f1 scores. It outperformed the baseline in all aspects by almost factor of two. For the average precision, the performance enhanced by a factor of five.

        \begin{table}[t]
        \begin{tabular}{|c||c|c|c|c|} \hline
                    & precision & recall    & f1    & average precision\\ \hline\hline
        Baseline	&0.310       &0.399	 	&0.349	&   0.144	\\ \hline
        Our method	&0.648     	 &0.775		&0.706  &   0.649	\\ \hline
        \end{tabular}
        \caption{\label{Tab1} Metrics for performance of baseline and our method.}
        \label{tab:metrics}
        \end{table}

    \subsection{DRL agent efficiency}

        We vary $M$ to analyze the efficiency of ROI proposal on the computation power. The intuition behind the different $M$ is that the DRL agent with a larger $M$ will develop a tendency to enclose more $p$ per $E$. The total computation power $C$ is dictated by the cost of running the fine detector and the DRL agent. The fine detector utilizes CNN architectures and primarily consumes the GPU power $C_{\text{GPU}}$. The fine detector is run at the end every episode, and $C_{\text{GPU}} \sim n_E$, where $n_E$ is the total number of $E$ in the detection task. The DRL agent in our method is mostly run on a CPU, $C_{\text{CPU}}$, due to the small size of neural network. The cost of operating the DRL agent is $C_{\text{CPU}} \sim n_t$, where $n_t$ is the average number of steps $t$ in $E$ during the entire detection task. As a result,
        \beq
            C \sim \frac{1}{\beta} n_E + \beta n_t,
            \la{eq:cost}
        \eeq
        
        \noindent where $\beta$ is determined by the cost ratio of one instance of fine detection and one $t$ iteration for the DRL agent. For instance, if the cost of taking a fine scan is much more expensive than updating the DRL agent for one step, $\beta \ll 1$. A na\"ive baseline for ROI proposal is one ROI ($E$) per $p$ that takes one $t$, which is $n_t=1$ and $n_E=N$. We compare this baseline and four different DRL agents, trained with different $M$ in Fig. \ref{fig:rl_efficiency}.

        Figure \ref{fig:rl_efficiency} (a) compares the four agents with different $M = 1,2,3,4$ based on the total number of episodes $n_E$. The agent with $M=1$ is nearly identical to the ROI baseline for all $N$. This indicates that the agent developed a policy, where it calls the stop action immediately upon securing one $p_\te$. In the very small $N$ limit ($N < 10$), all agents are similar to the baseline performance. In this regime, $p$ are so sparsely located that the different policies of DRL agents do not lead to difference in $n_E$. However, when $N>10$, the agents with $M=2,3,4$ start to deviate from the baseline and $M=1$. The difference in performance between the $M=2$ and $M=3,4$ agents become more apparent after $N>100$. The agents $M=3,4$ remain similar throughout all $N$. One possible explanation for the similarity between $M=3$ and $M=4$ is that $b$ rarely encloses more than 3 $p$ because the dimensions of $b$ is not large enough. As a result, the $M=4$ agent converges to the a similar policy as the $M=3$ agent.
        
        Figure \ref{fig:rl_efficiency} (b) reveals the performances of the four agents $M=1,2,3,4$ on $n_t$ per $E$. All agents show decreasing $n_t$ with the increasing $N$; it takes less $n_t$ for a DRL agent to reach the nearest $p$. Again, the $M=3,4$ agents exhibit similar results, implying that they converged to the similar policies. 
        
        For the present experiment and other searches that are similar, we conclude that either of $M=3, 4$ agents is the optimal solution. When $N$ is small ($N<100$), the $M=3$ agent is similar to other agents and better than the baseline in $n_E$. For large $N$ ($N>100$), it outperforms other agents in $n_E$. The agents $M=1,2$ are better in $n_t$. However, in our setting, $C_{\text{GPU}} \gg C_{\text{CPU}}$ and $\beta \ll 1$ in \eqref{eq:cost}. Therefore, the advantage in $n_E$ of the $M=3$ agent compensates for the inefficiency in $n_t$. The agent $M=4$ shows almost identical performance as the $M=3$ agent does, and the training time was also similar \cite{sm}.
        
\section{Conclusion}

    We propose a method to efficiently and accurately locate and identify target crystals from a high resolution image, which contains a large number of microflakes that differ immensely in their sizes and shapes. The searching process is based on deep reinforcement learning and object detection algorithms. It is divided into two stages, which are the coarse and fine search. During the coarse search, a DRL agent proposes the ROI based on the highly compressed and downsampled version of original image. A fine detector in the fine search scrutinizes the ROI in the high resolution image.

    Our method takes advantage of the bimodal color distribution of exfoliated crystals on a substrate. The resulting RL framework is based on states instead of pixels. Its state and action spaces are small and independent of the fine detector. The DRL agent requires a simple neural network, converges quickly during training, and is computationally inexpensive to iterate. Our method is a high-level design that can be tailored for detecting a variety of crystals.

    We experimented our method on searching target $\moly$ crystals for edge supercurrent measurements. Metrics (AP and f1 score) show that our method surpassed the baseline in performance. We analyze the efficiency of DRL agents in computing power and show that either of the agents trained with $M=3, 4$ is optimal for the present experiment.

\acknowledgments
S.K. thanks Jee Kim and Jane Cha for their help on preparing the dataset.


\onecolumngrid

\bfig
    \ing[width=\columnwidth]{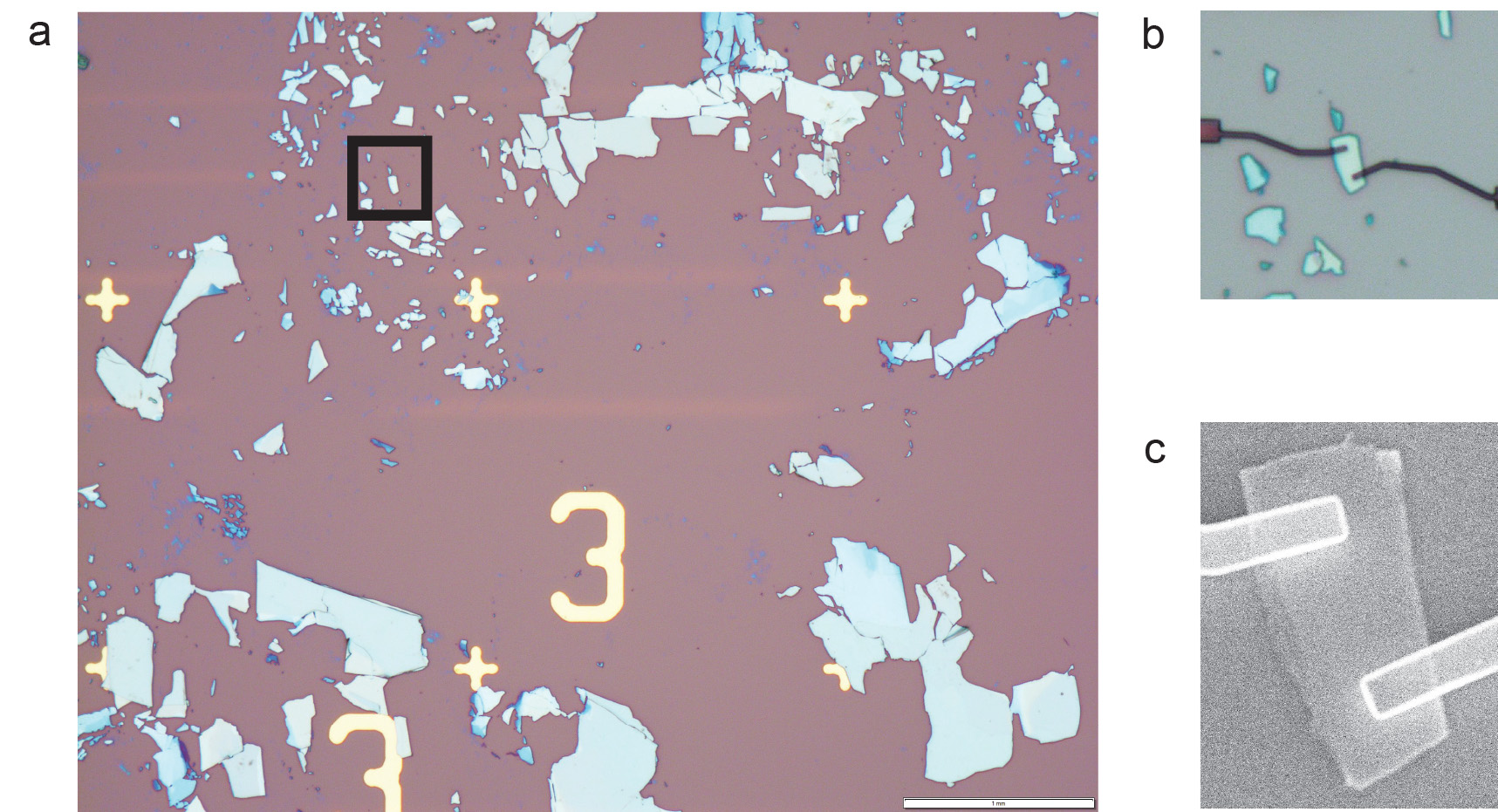}
    \caption{Images of crystals. (a) A typical substrate after a 2D material is exfoliated. It contains a large number of crystals with different shapes, thicknesses, and sizes. (b) A zoomed-in image of the region enclosed by the black box in (a). A target crystal is located in the center. This image is taken after a process in nanofabrication, and the target crystal has electrode patterns. (c) A scanning electron microscopy (SEM) image of the target crystal from Panel (b). The target crystal now has evaporated electrodes on it.}
    \label{fig:device_images}
\efig

\bfig
    \ing[width=\columnwidth]{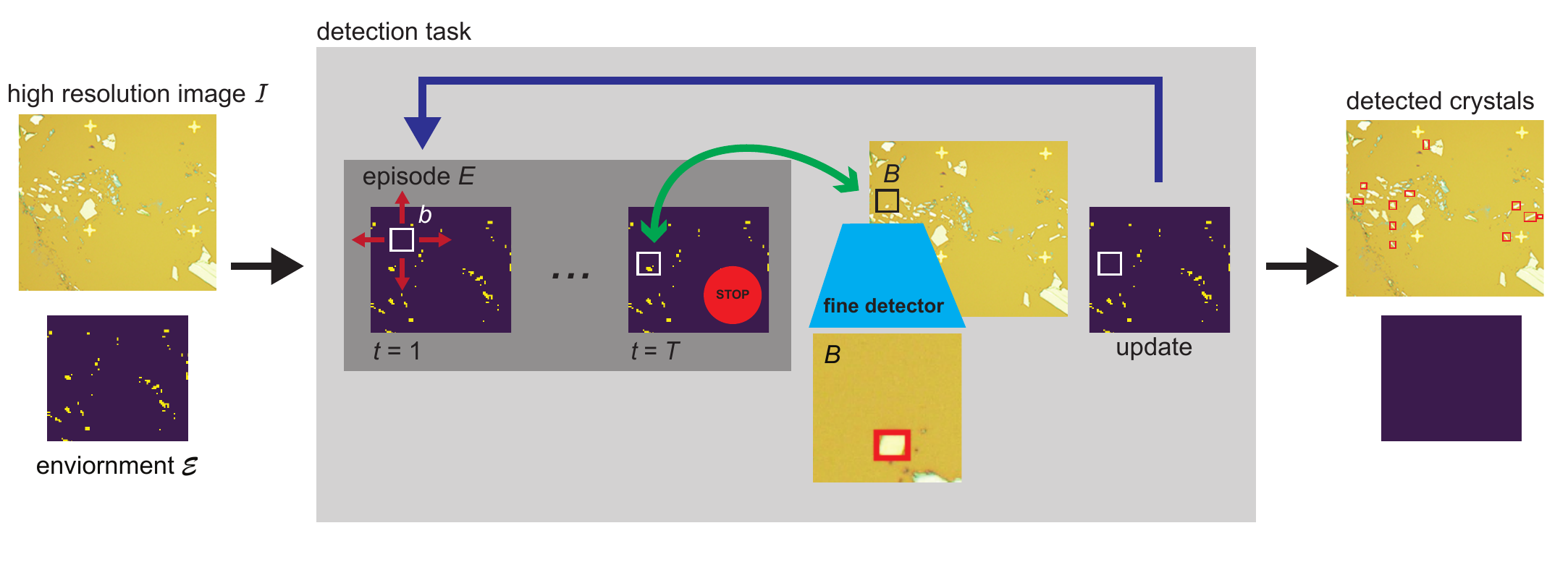}
    \caption{The workflow of our method. A high resolution image $\cI$ and the corresponding RL environment $\cE$ are prepared. The iterative ROI proposals and fine detector search polygons and crystals. After the entire detection task is completed, the locations of crystals in $\cI$ are marked with bounding boxes. All $\{p\}$ are removed from $\cE$.}
    \label{fig:workflow}
\efig

\bfig
    \ing[width=\columnwidth]{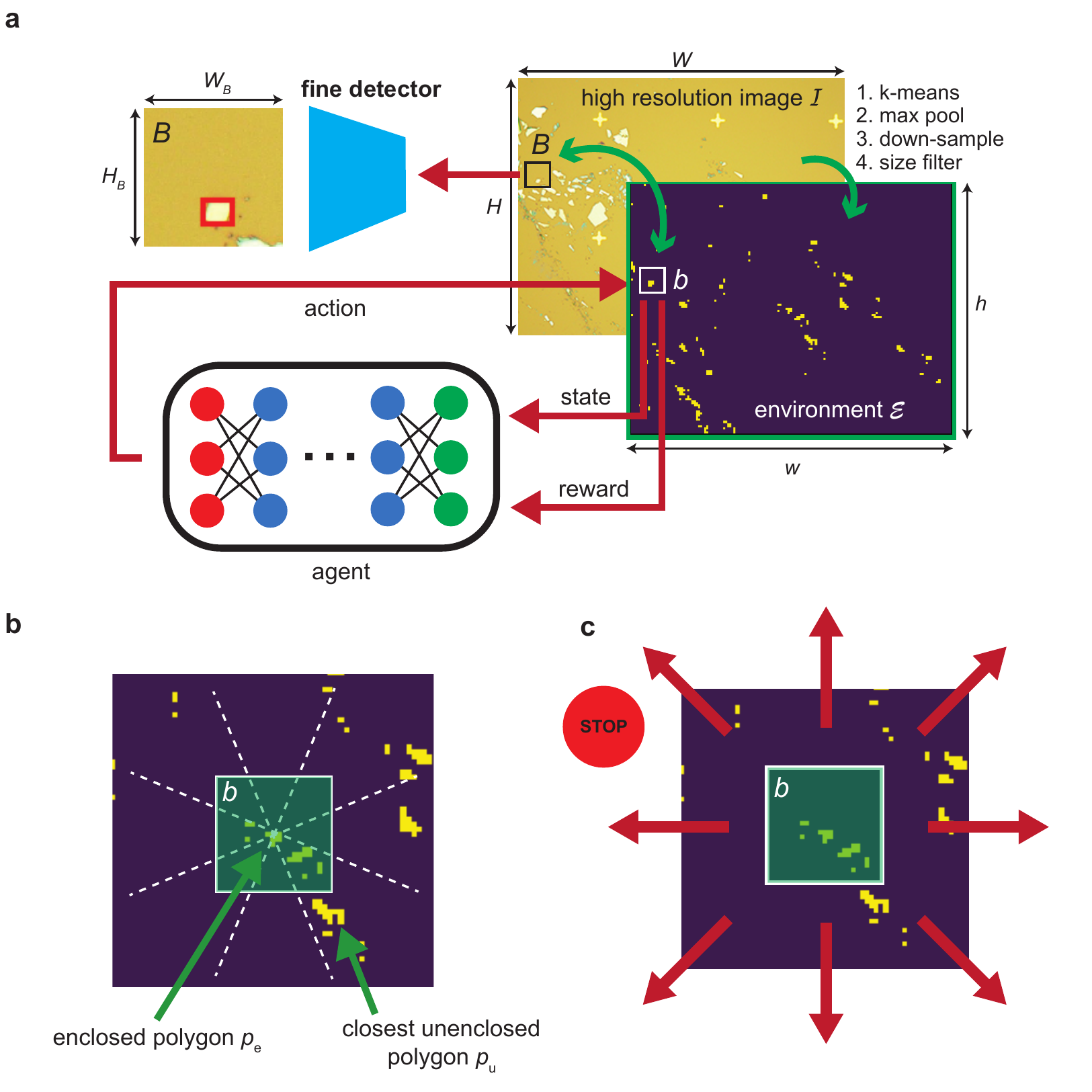}
    \caption{(a) A cartoon of DRL agent, fine detector, $\cI$, and $\cE$ in our method. (b) The segmented surroundings near $b$. The dotted line represents the boundaries between different sections. Each section in $b$ (colored green) corresponds to an element in $\vtheta_\te$ and that of outside to an element in $\vtheta_\tu$. (c) The action space $\cA$ of DRL agent in our method.}
    \label{fig:architecture}
\efig

\bfig
    \ing[width=\columnwidth]{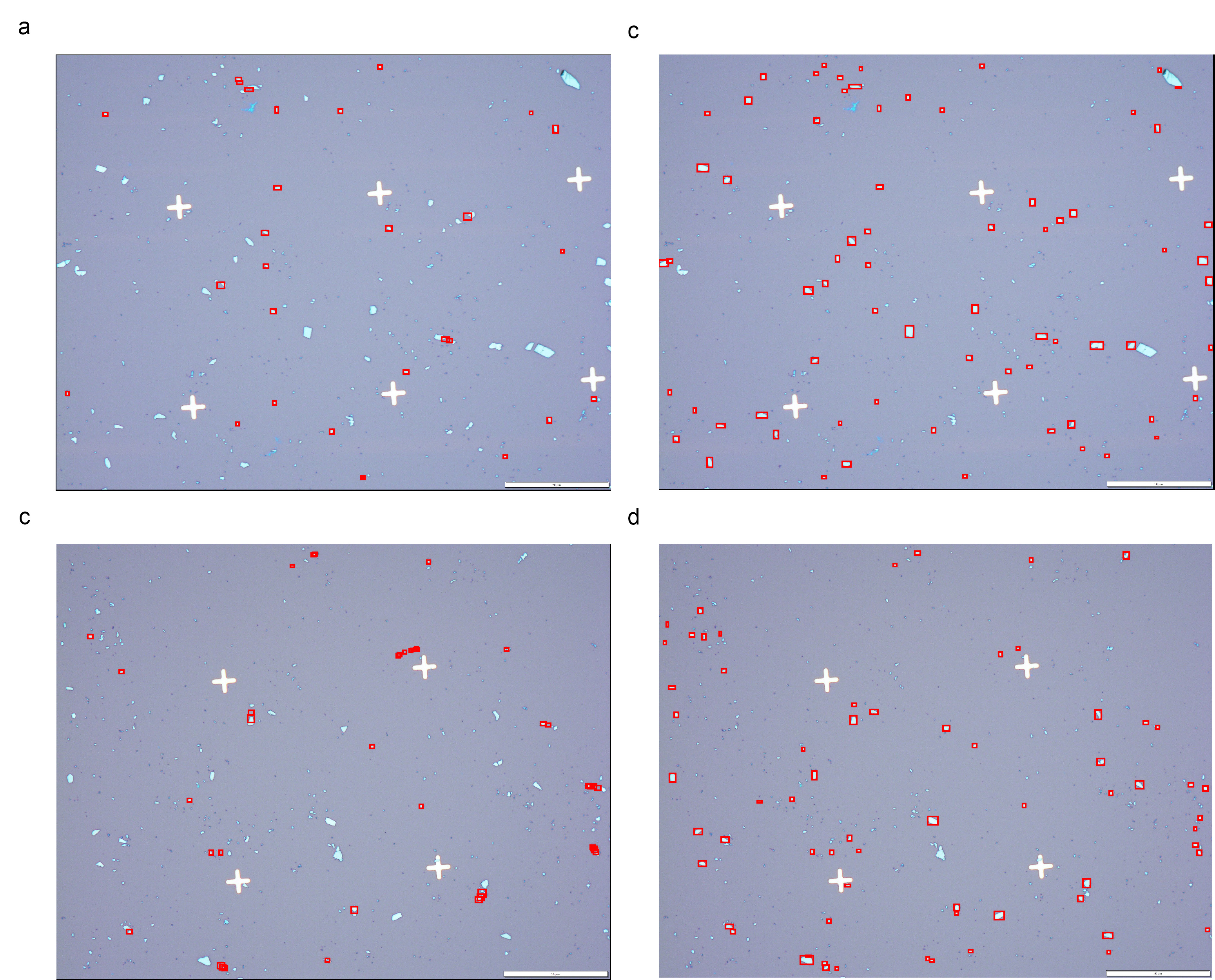}
    \caption{Comparison of detection performance of the na\"ive baseline (Panels a and c) and our method (Panels b and d). The red boxes are the bounding boxes that the corresponding algorithm drew around the predicted crystals.}
    \label{fig:qualitative_analysis}
\efig

\bfig
    \ing[width=\columnwidth]{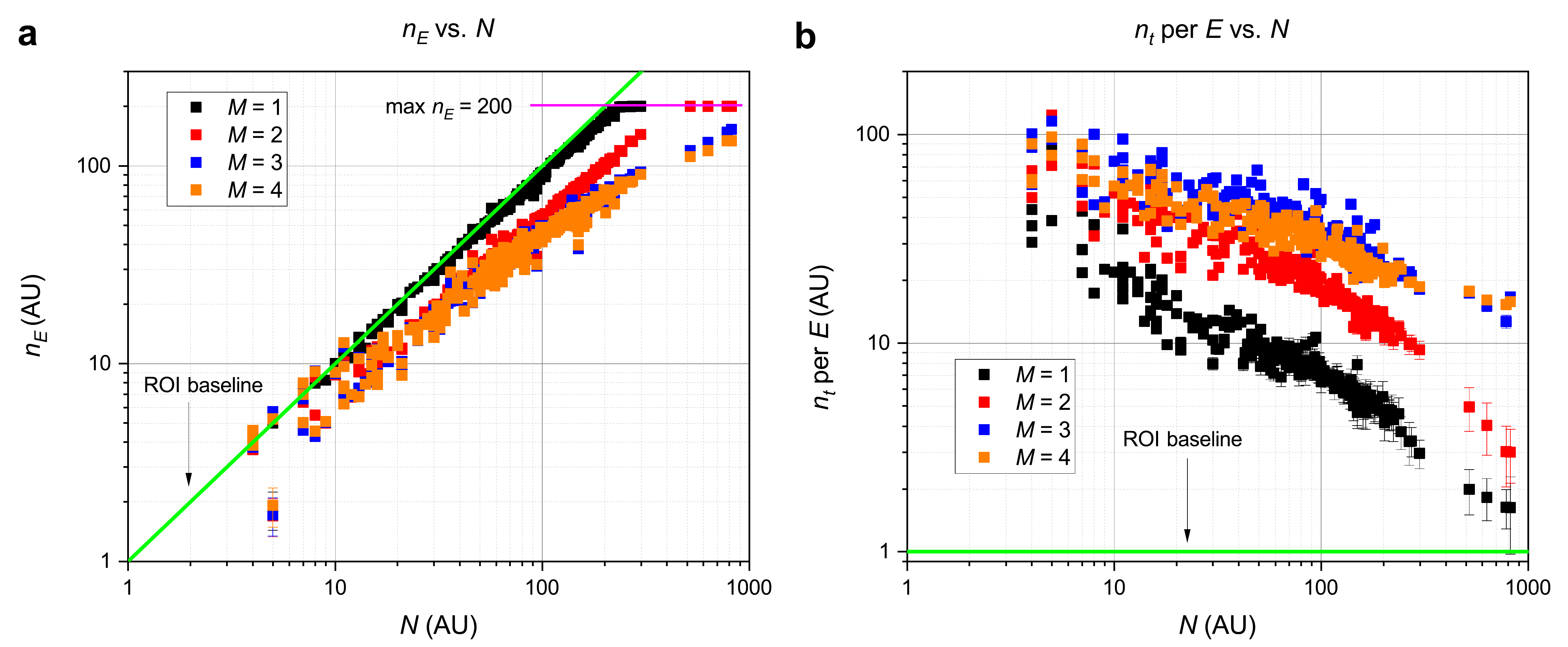}
    \caption{Comparison of DRL agents trained with different $M = \{1,2,3,4\}$. The maximum number of episodes during evaluation was set to 200. Each dot represents evaluation of a DRL agent on $\cE$, created from $\cI$. (a) The number of episodes $n_E$ to complete the entire detection task versus the total number of polygons $N$ in $\cE$. (b) The number of average steps $n_t$ per $E$ versus $N$.}
    \label{fig:rl_efficiency}
\efig

\end{document}